\documentclass[12pt]{article}
\usepackage{amsfonts}
\usepackage{amsmath}
\usepackage{amssymb}
\usepackage{amstext}
\usepackage{amsfonts}
\usepackage{graphicx,epsfig}
\usepackage{indentfirst}
\usepackage{hyperref}

\newtheorem{theorem}{Theorem}
\newtheorem{proposition}[theorem]{Proposition}
\makeatletter
\def\ExtendSymbol#1#2#3#4#5{\ext@arrow 0099{\arrowfill@#1#2#3}{#4}{#5}}
\def\RightExtendSymbol#1#2#3#4#5{\ext@arrow 0359{\arrowfill@#1#2#3}{#4}{#5}}
\def\LeftExtendSymbol#1#2#3#4#5{\ext@arrow 6095{\arrowfill@#1#2#3}{#4}{#5}}
\makeatother

\begin{document}
\baselineskip 20pt

\title{Local Unitary Classification of Arbitrary Dimensional Multipartite Pure States}
\author{Bin Liu$^1$, Jun-Li Li$^1$, Xikun Li$^1$, and
Cong-Feng Qiao$^{1,2}$\footnote{Corresponding author.}\\[0.5cm]
$^{1}$Department of Physics, Graduate University of Chinese Academy of Sciences \\
YuQuan Road 19A, Beijing 100049, China\\[0.2cm]
$^{2}$Theoretical Physics Center for Science Facilities (TPCSF),
CAS\\ YuQuan Road 19B, Beijing 100049, China}

\date{}
\maketitle

\begin{abstract}
We propose a practical entanglement classification scheme for
general multipartite pure states in arbitrary dimensions under local
unitary equivalence by exploiting the high order singular value
decomposition technique and local symmetries of the states. By
virtue of this scheme, the method of determining the local unitary
equivalence of $n$-qubit states proposed by Kraus is extended to the
case for arbitrary dimensional multipartite states.\\


\end{abstract}

Entanglement is one of the most extraordinary features of quantum
theory. It lies at the heart of quantum information theory and is
now regarded as a key physical resource in realizing many quantum
information tasks, such as teleportation and quantum computation,
etc. \cite{quantum-computation}. In practice, people may confront
various forms of entangled states even in one single physical
system. Though superficially showing up with different features,
usually not all these entangled states are functionally independent;
they may be intrinsically the same where the entanglement property
is concerned. Two entangled states are said to be equivalent in
implementing the same quantum information task if they can be
obtained with certainty from each other via local operation and
classical communication (LOCC). Theoretically, this LOCC equivalent
class is defined such that within this class any two quantum states
are interconvertible by local unitary (LU) operators
\cite{three-qubit}. If we release the constraint of the local
unitary operator to invertible local transformation, the LOCC
equivalence then turns to the widely discussed stochastic local
operation and classical communication (SLOCC) equivalence.

The classification of a bipartite system under LU can be done by
singular value (Schmidt) decomposition. For the classification of a
pure multipartite entangled state, a canonical method was proposed
in Ref.\cite{Carteret-Sudbery}, though it was given only in a set of
constraints on the coefficients of quantum state. Later, this method
was reformulated into a compact form \cite{Verstraete-Moor}, where
the existence of SLOCC normal forms was also proved. Recently, by
introducing the standard form for multipartite states, Kraus
proposed a general way to determine the local unitary transformation
between two LU equivalent $n$-qubit states
\cite{Kraus-PRL104-020504}. Even though these beautiful results have
been obtained, a generic method or methodology is still missing in
the classification of general multipartite entangled states
\cite{R.H-P.H-M.H-K.H}.

Much research indicates that symmetry study is very helpful in the
classification of a general multipartite system. It is broadly
realized that making restrictions on the quantum states, i.e.,
satisfying some symmetric properties, is a feasible way in the
entanglement classification \cite{Tsubasa-Nobuhiro,
SLOCC-symmetry-n}. By virtue of the Majorana representation, a lot
of effort was spend on the study of the relation between permutation
symmetry and the classification of a multipartite entangled state
under both LOCC and SLOCC \cite{Markham,Ribeiro-Mosseri,Aulbach}. It
was recently realized that there exists one kind of local (internal)
symmetry under SLOCC equivalence for a continuously entangled state
\cite{parametric}. Unlike the permutation symmetry among different
partites, the internal symmetry is of a localized one. In
Refs.\cite{Lyons-Walck,Curt-Walck,Curt-Lyons}, people investigated
the nature of the $n$-qubit entangled state via analyzing the
stabilizers of the symmetric state, which can also be regarded as a
local unitary transformation.

In this work, we present a new and easy to operate entanglement
classification method under the LU equivalence, by virtue of the
technique of high order singular decomposition (HOSVD)
\cite{multi-singular-decom, Tensor-decomp} and by exploiting the
local symmetries of the state \cite{parametric}. The procedure of
this method goes as follows: Express the quantum state into a
complex tensor, analyze the possible local symmetry property of the
quantum state, apply the HOSVD on the tensor form of the quantum
state, and then fully classify a general multipartite state into LU
inequivalent classes by sequential use of HOSVD and local unitary
symmetry.

A general $N$-partite entangled quantum state in dimension
$I_1\times I_2 \times \cdots \times I_N$ can be formulated as the
following form:
\begin{eqnarray}
|\Psi\rangle=\sum_{i_1=1,i_2=1,...,i_N=1}^{I_1,I_2,\cdots,I_N}
\psi_{i_{1}i_{2}...i_{N}} |i_1\rangle|i_2\rangle...|i_N\rangle \; ,
\end{eqnarray}
where $\psi_{i_{1}i_{2}...i_{N}}\in \mathbb{C}$ are coefficients of
the quantum state in representative bases. Two quantum states are
said to be equivalent if they are interconvertible by a certain type
of operators, which can be schematically expressed as
\begin{eqnarray}
|\Psi'\rangle & = & \bigotimes_{i}^{N} U^{(i)} |\Psi\rangle
\nonumber
\\ & = &
\sum_{\substack{i_{1},i_{2},\cdots,i_{N} \\
i'_{1},i'_{2},\cdots,i'_{N}}} \psi_{i_{1}i_{2}\cdots i_{N}}
u^{(1)}_{i'_{1}i_1}|{i'_{1}\rangle u^{(2)}_{i'_{2}i_2}|i'_{2}\rangle
\cdots u^{(N)}_{i'_{N}i_N}|i'_{N}}\rangle \; \nonumber \\
& = & \sum_{i_{1},i_{2},\cdots,i_{N}} \psi'_{i_{1}i_{2}\cdots i_{N}}
|{i_{1},i_{2},\cdots,i_{N}}\rangle \; . \label{psi-equiv-coe}
\end{eqnarray}
Here, the coefficients $\psi_{i_{1}i_{2}...i_{N}}$ can also be
treated as the entries of a tensor $\Psi$, and hence the quantum
states can be represented by high dimensional complex tensors. From
the tensor form of $\Psi$, the operator $U^{(n)}$ acting on the
$n$th partite is defined as
\begin{eqnarray}
(U^{(n)}\Psi )_{i_1i_2\cdots i_{n-1}i_n'i_{n+1\cdots i_N}} \equiv
\sum_{i_n} \psi_{i_1i_2\cdots i_{n-1}i_ni_{n+1}\cdots i_N}
u^{(n)}_{i'_{n}i_n} \; .
\end{eqnarray}
In case $U^{(n)}$'s are unitary operators, quantum state $\Psi'$ is
said to be LU equivalent to $\Psi$, i.e.,
\begin{eqnarray}
\Psi'=\bigotimes_{i} U^{(i)} \Psi \; . \label{A-def}
\end{eqnarray}

Suppose a concerned quantum state is inherited with local
symmetries, like
\begin{eqnarray}
\bigotimes_n P^{(n)} \Psi = \Psi \; ,
\end{eqnarray}
and is always assisted with classical communication hereafter in our
discussion; the local symmetry group $G=\{g|
g=\bigotimes_{n}P^{(n)}\}$ then forms the stabilizer of $\Psi$. Here
$P^{(n)}$ is the operator that acts on the $n$th partite. From
Eq.(\ref{A-def}), we can get
\begin{eqnarray}
\Psi' & = &  U^{(1)}\otimes U^{(2)}\otimes
\cdots \otimes U^{(i)} \otimes \cdots\otimes U^{(N)} \Psi \nonumber \\
& = & U^{(1)}P^{(1)}\otimes U^{(2)}P^{(2)}\otimes \cdots
\nonumber \\ & & \otimes U^{(i)}P^{(i)} \otimes \cdots
\otimes U^{(N)}P^{(N)} \Psi \nonumber \\
& = &  \bigotimes_{i}U^{(i)}P^{(i)}U^{(i)\dag} \Psi'\;
\end{eqnarray}
with $U^{(i)}$ the unitary operators. Clearly, every local unitary
equivalent state $\Psi'$ has the local symmetry of
$(\bigotimes_{i}U^{(i)}) \cdot (\bigotimes_{j}P^{(j)})
\cdot(\bigotimes_{k}U^{(k)\dag})$, which is isomorphic to that of
$\Psi$. Consequently, we have the following proposition.
\begin{proposition}
If $\Psi'$ is local unitary equivalent to $\Psi$, then the
stabilizers of the local symmetries of the two states are unitarily
equivalent.
\end{proposition}

By taking $|\psi\rangle = \alpha|11\rangle +\beta|22\rangle +
\gamma|33\rangle +\gamma|44\rangle$, where
$|\alpha|^2+|\beta|^2+2|\gamma|^2 = 1$ as an example, this $4\times
4$ entangled state can be expressed in a matrix form:
\begin{eqnarray}
\Psi_{4\times 4} = \begin{pmatrix} \alpha & 0 & 0 & 0\\ 0
& \beta & 0 & 0 \\
0 & 0 & \gamma & 0 \\  0 & 0 & 0 & \gamma
\end{pmatrix} \; . \label{44singular}
\end{eqnarray}
It is invariant under the local transformation $U^{(1)} =
\{e^{i\theta_1} , e^{i\theta_2}, u\}$ and $U^{(2)} =
\{e^{-i\theta_1} , e^{-i\theta_2}, u^{*}\}$, i.e.,
\begin{eqnarray}
U^{(1)}\cdot\Psi\cdot (U^{(2)})^{\mathrm{T}} = \Psi \; .
\end{eqnarray}
The $\Psi_{4\times 4}$ is then a $U^{(1)}\otimes U^{(2)}$ invariant
quantum state.

To study the LU classification of a multipartite system, it is
convenient for us to use the notation of
Ref.\cite{multi-singular-decom} and to define the matrix unfolding
of the tensor $\Psi\in \mathbb{C}^{I_1I_2 \cdots I_{N}} $ with the
$n$th index as
\begin{eqnarray}
\Psi_{(n)} \in \mathbb{C}^{I_n\times (I_{n+1}I_{n+2}\cdots
I_{N}I_1I_2\cdots I_{n-1})} \; . \label{matrix-unfolding-def}
\end{eqnarray}
Here $\Psi_{(n)}$ is an $I_n\times (I_{n+1}I_{n+2}\cdots
I_{N}I_1I_2\cdots I_{n-1})$ matrix. Considering the high order
singular value decomposition developed in Ref.
\cite{multi-singular-decom}, there exists a core tensor $\Sigma$ for
each $\Psi$, that is,
\begin{eqnarray}
\Psi = U^{(1)}\otimes U^{(2)} \cdots \otimes U^{(N)} \Sigma\; ,
\end{eqnarray}
where $\Sigma$ forms a same order tensor in the Hilbert space
$I_1\times I_2\times \cdots \times I_N$. Any $N$-$1$ order tensor
$\Sigma_{i_n=i}$, obtained by fixing the $n$th index to $i$, has the
following properties \cite{multi-singular-decom}:
\begin{eqnarray}
\langle \Sigma_{i_n=i}, \Sigma_{i_n=j} \rangle =
\delta_{\scriptstyle ij}\,\sigma_{i}^{(n)\,2}\; ,
\end{eqnarray}
with $ \sigma_{i}^{(n)} \geq \sigma_j^{(n)}$ and $\forall\; i<j$ for
all possible values of $n$. Here, the singular value
$\sigma_i^{(n)}$ symbolizes the Frobenius norm $\sigma_{i}^{(n)} =
||\Sigma_{i_n=i}|| \equiv
\sqrt{\langle\Sigma_{i_n=i},\Sigma_{i_n=i}\rangle}$, where the inner
product $\langle \mathcal {A}, \mathcal {B} \rangle \equiv
\sum_{i_{1}} \sum_{i_{2}} \cdot\cdot\cdot \sum_{i_{N}}
b_{i_1i_2...i_N} a^*_{i_1i_2...i_N}$.

From the definition, we know that if the quantum state $\Psi'$ is LU
equivalent to $\Psi$, then they can be transformed into the same
core tensor $\Sigma~$. However, because there is no information on
whether different core tensors are local unitary equivalent or not,
$\Sigma$ has nothing to do with the entanglement classification of
the multipartite state yet. Hence, to achieve the entanglement
classification from this point, the following analysis is necessary.

Since every unitarily equivalent quantum state possesses the same
local symmetric property, we can pick up a typical state to
represent one equivalent class. A natural choice is the core tensor
for each class. In the matrix unfolding form of the core tensor,
$\Sigma_{(n)}$ can be expressed as
\begin{eqnarray}
\Psi_{(n)} = U^{(n)} \cdot \Sigma_{(n)} \cdot ( U^{(n+1)}\otimes
\cdots \otimes U^{(N)} \cdot U^{(1)}\cdots U^{(n-1)})^{\mathrm{T}}
\; , \label{unfolding-Psi-n}
\end{eqnarray}
where $U^{(n)}$ is constructed by the left singular vectors of
$\Psi_{(n)}$.

There exist such quantum states $\Psi$ that some of the singular
values of their unfolding matrix $\Psi_{(n)}$ are identical, which
is similar to the singular value $\gamma$ in the matrix case in
Eq.(\ref{44singular}) and is also stated in Property 4 of
Ref.\cite{multi-singular-decom}. The $\Psi_{(n)}$ has the following
property:
\begin{eqnarray}
\Psi_{(n)} = U^{(n)}P^{(n)} \cdot  \Sigma_{(n)}' \cdot (
U^{(n+1)}\otimes \cdots \otimes U^{(N)} \otimes U^{(1)}\cdots
U^{(n-1)})^{\mathrm{T}} \; ,
\end{eqnarray}
where $P^{(n)}$ is a block-diagonal matrix consisting of unitary
blocks with the same partitions as that of the identical singular
values, and $\Sigma_{(n)}' = P^{(n)\dag}\Sigma_{(n)}$. For such
states we have
\begin{eqnarray}
\Sigma' = \bigotimes_{n=1}^{N}P^{(n)\dag}\Sigma \; .
\label{local-symmetry-core}
\end{eqnarray}
Here, $P^{(n)}= \text{diag} \{e^{i\theta_1}, \cdots,
e^{i\theta_{I_n}}\}$ in case there is no identical singular value
for $\Sigma_{(n)}$. Thus, if $\Sigma$ is a HOSVD of $\Psi$, then the
$\Sigma'$ is as well. In other words, the HOSVD of $\Psi$ possesses
local symmetry properties of $\bigotimes_{n=1}^{N}P^{(n)\dag}$.

From the above discussion, we may conclude that if we introduce the
symmetry property to compensate the nonuniqueness of the core
tensor, it will then serve as the unique representative state for
each local unitary equivalent class. We call the representative
state $\Sigma$ associated with specific symmetry group
$\bigotimes_{n=1}^{N}P^{(n)}$ the unique canonical form of the
entanglement class under LU equivalence. Two quantum states are LU
equivalent if and only if they have the same core tensor
(representative state) or their core tensors are related by the
symmetry group $\bigotimes_{n=1}^{N}P^{(n)}$.
\begin{proposition}
The high order singular value decomposition of multipartite state is
a LU classification of entanglement up to the local symmetry
$\bigotimes_{n=1}^{N}P^{(n)}$.
\end{proposition}
Proof: First, it is easy to see that if two quantum state are LU
equivalent, i.e., $\Psi'=\otimes_{i}^{N} U^{(i)} \Psi$, then they
can be transformed into the same core tensor $\Sigma$. Second,
suppose two core tensors are related by local unitary
transformations $\Sigma'=\bigotimes_iV^{(n)}\Sigma$, where
$\bigotimes_iV^{(n)}$ are different from
$\bigotimes_{n=1}^{N}P^{(n)}$; then by taking the $n$th unfolding as
an example, we have
\begin{eqnarray}
\Psi'_{(n)} = U^{(n)} \cdot \Sigma'_{(n)} \cdot (\cdots)  = U^{(n)}
V^{(n)} \cdot\Sigma_{(n)}\cdot (\cdots)\; .
\end{eqnarray}
This means that the singular values are not uniquely determined,
which is in contradiction with Property 4 of Ref.
\cite{multi-singular-decom}. Therefore, we can conclude that the
core tensor $\Sigma$ up to the local symmetry
$\bigotimes_{n=1}^{N}P^{(n)}$ is unique, or, in other words, the
core tensor of the quantum state can be treated as the LU
classification of a multipartite quantum state up to the local
symmetry $\bigotimes_{n=1}^{N}P^{(n)}$. QED.

Now, for a general arbitrary dimensional $N$-partite pure state, the
canonical form can be constructed in the following way. First,
express the tensor form quantum state in matrix unfolding form
$\Psi_{(n)}$ [see Eq.(\ref{matrix-unfolding-def})]. Then, compute
the singular value decomposition of the matrix $\Psi_{(n)} =
U^{(n)}\Lambda V^{(n)}$. Here $N$ unitary matrices $U^{(n)}$,
composed by left singular vectors of $\Psi_{(n)}$, for $N$ indices
will be obtained, and this is a simple process while only matrix
operations are involved. Finally, the core tensor is constructed by
\begin{eqnarray}
\Sigma = \bigotimes_{n=1}^{N} U^{(n)\dag} \Psi \; .
\end{eqnarray}
The representative state $\Sigma$ with its associated symmetry group
in the form of Eq.(\ref{local-symmetry-core}) will serve as the
unique canonical form of $\Psi$.

Given two canonical forms, i.e., the obtained representative states
$\Sigma$ and $\widetilde{\Sigma}$ with some local symmetries of
$\bigotimes_n Q^{(n)}$ and $\bigotimes_n \widetilde{Q}^{(n)}$,
respectively, the procedure of verifying the existence of a local
symmetry that connects the two representative states is
straightforward. If singular values of these two representative
states or the sizes of the unitary blocks of local symmetries are
not the same, obviously there does not exist such a local symmetry
that the two tensors can be transformed into each other. Otherwise,
the existence of the local symmetry is equivalent to the existence
of solutions to the equation $\widetilde{\Sigma} = \bigotimes_n
Q^{(n)} \Sigma$ (or $ \Sigma = \bigotimes_n \widetilde{Q}^{(n)}
\widetilde{\Sigma}$), where the elements in $\Sigma$ and
$\widetilde{\Sigma}$ are known quantities whereas they are unknown
variables in $Q^{(n)}$ ($\widetilde{Q}^{(n)}$). By virtue of the
mathematical technique {\it relinearization} \cite{Kipnis-Shamir,
Courtois-Klimov-Patarin-Shamir}, one can readily know whether the
solutions to the equations exist or not (practical examples are
given in Ref. \cite{LU-Exmaple}).

Last, we show briefly the relation of the core tensor and Kraus's
canonical form $|\Psi_{\mathrm{st}}\rangle$ in Ref.
\cite{Kraus-PRL104-020504}. The sorted trace decomposition of the
$n$th partite requires $\mathrm{Tr}_{\neg\,
i_n}(|\Psi_{\mathrm{st}}\rangle\langle \Psi_{\mathrm{st}}|)$ to be
diagonalized. For a general quantum state
\begin{eqnarray}
|\Psi_{\mathrm{st}}\rangle & = &
\sum_{i_{1},i_{2},\cdots,i_{N}}^{I_1,I_2,\cdots I_N}
\psi_{i_{1}i_{2}\cdots i_{N}} |{i_{1},i_{2},\cdots,i_{N}}\rangle
\nonumber \\ & = & \sum_{i_n}|i_n\rangle \sum_{\neg i_n}
\psi_{i_{1}i_{2}\cdots i_{n-1}i_n i_{n+1}\cdots i_{N}}
|{i_{1}i_{2}\cdots i_{n-1}i_{n+1}\cdots i_{N}}\rangle \; ,
\end{eqnarray}
the requirement of trace diagonalization asks
\begin{eqnarray}
\sum_{\neg j} \psi^*_{i_{1}i_{2}\cdots i_{n-1}j i_{n+1}\cdots i_{N}}
\psi_{i_{1}i_{2}\cdots i_{n-1}j' i_{n+1}\cdots i_{N}} =
\lambda_j^{(n)}\delta_{jj'} \; ,\nonumber \\ \lambda_1^{(n)} \geq
\lambda_2^{(n)} \geq \cdots \geq \lambda_{I_{n}}^{(n)} \geq 0 \; .
\end{eqnarray}
Here, $\lambda_j^{(n)} = \sum_{\neg j} a^*_{i_{1}i_{2}\cdots
i_{n-1}j i_{n+1}\cdots i_{N}} a_{i_{1}i_{2}\cdots i_{n-1}j
i_{n+1}\cdots i_{N}} $  is just the Frobenius norm
$||(\Psi_{\mathrm{st}})_{i_n=j}||^2$ in high order singular
decomposition employed in our classification approach.

In summary, by virtue of the symmetry property study and tensor
decomposition we propose an effective and easy to operate method for
the local unitary classification of a multipartite entangled state,
which splits the mathematically difficult problem into different
relatively easier ones. That is, we first play tensor decomposition
on the quantum state and obtain the quasinormal form (core tensor),
where many different quasinormal forms may actually belong to the
same entanglement class; then, by exploiting the local symmetries
between the quasinormal forms, we arrive at a uniquely defined
entanglement class. By this approach, the verification of LU
equivalence of $n$-qubit states proposed by Kraus is extended to the
case for arbitrary dimensional multipartite states in a more easily
understandable and computable algorithm. From a methodological point
of view, the approach in this work is different from previous ones
in the literature, where normally the entangled state is first
divided into apparently inequivalent coarse-grained sets under
certain criteria (like range \cite{Range-1}, partition number
\cite{SLOCC-symmetry-n}, ranks \cite{2nn}, etc.) and then more
fine-grained classification procedures are performed. This new
approach might also be useful in the study of classification under
SLOCC or other characters of the general multipartite entangled
state. Finally, it should be mentioned that the tensor decomposition
method may also play an important role in other entanglement
property studies \cite{Joseph-Sudbery}.

\vspace{0.7cm} {\bf Acknowledgments}

This work was supported in part by the National Natural Science
Foundation of China(NSFC), by the CAS Key Projects KJCX2-yw-N29 and
H92A0200S2.

\end{document}